# Axial vs. Radial Junction Nanowire Solar Cell

Vidur Raj[*], Hark Hoe Tan[*] and Chennupati Jagadish[*]

[1]Department of Electronic Materials Engineering, Research School of Physics, The Australian National University, Canberra, ACT 2601, Australia

vidur.raj@anu.edu.au; Hoe.Tan@anu.edu.au; Chennupati.Jagadish@anu.edu.au

**Abstract**

Both axial and radial junction nanowire solar cells have their challenges and advantages. However, so far, there is no review that explicitly provides a detailed comparative analysis of both axial and radial junction solar cells. This article reviews some of the recent results on axial and radial junction nanowire solar cells with an attempt to perform a comparative study between the optical and device behavior of these cells. In particular, we start by reviewing different results on how the absorption can be tuned in axial and radial junction solar cells. We also discuss results on some of the critical device concepts that are required to achieve high efficiency in axial and radial junction solar cells. We include a section on new device concepts that can be realized in nanowire structures. Finally, we conclude this review by discussing a few of the standing challenges of nanowire solar cells.

**Introduction**

It has been predicted that nanowires can significantly reduce the overall cost of solar cells because of several of their advantages compared to their thin-film counterpart. Some of these advantages include but are not limited to: (a) high generation rate per unit volume of material used, (b) low-cost and fast growth rate compared to their thin-film counterpart, (c) facile strain relaxation allows for heteroepitaxy, and (d) increased defect tolerance [1-14]. Furthermore, from the device point of view, nanowire architecture allows for the fabrication of new and innovative device structures, which is either very complicated or not possible in a thin film solar cell [15, 16]. Moreover, it also provides us with the possibility of fabricating low cost, high performance, lightweight, highly flexible solar cells [8-14, 17].

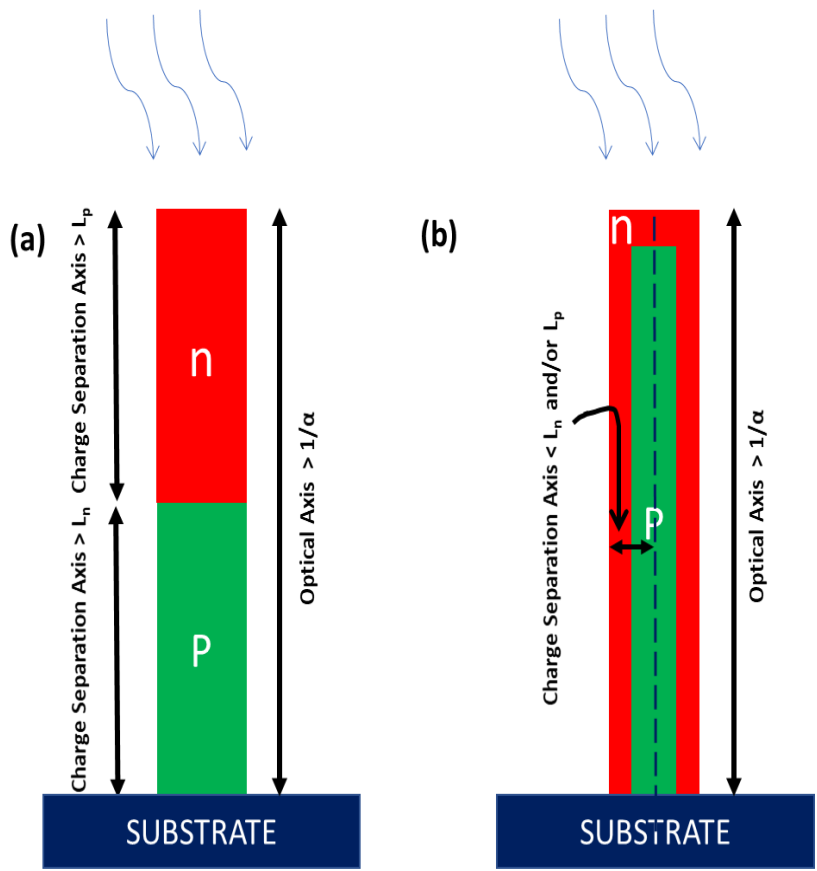

**Figure 1.** Schematic representation of (a) an axial p-n junction nanowire solar cell and (b) a radial p-n junction nanowire solar cell. In both of these figures, α denotes the absorption coefficient of the active material and $L_n$ and $L_p$, denotes the electrons and holes diffusion lengths, respectively. *[Figures 1(a)-(c) have been reprinted (adapted) with permission from ref [15] Copyright (2019) IEEE.]*

A nanowire solar cell can be broadly classified into an axial or a radial junction device based on the axis of charge carrier separation. In an axial junction solar cell, the charge carrier separation happens along the length of the nanowire, whereas in a radial junction solar cell, charge carrier separation happens along the radial axis. Technically a radial junction is only required when the bulk lifetime of the material is very low [15, 16]. Figures 1(a) and 1(b) show light absorption and charge carrier separation in an axial and a radial junction nanowire solar cell. In a solar cell, the minimum length required to achieve sufficient absorption is characterized by absorption depth ($1/\alpha$), which describes how deeply light penetrates the semiconductor before being absorbed. At the same time, diffusion length characterizes the maximum length that the minority carrier can travel before recombining non-radiatively. For efficient operation of a solar cell, the diffusion length should be higher than the absorption depth, as schematically depicted in Figure 1. In cases where the minority carrier lifetime (or diffusion length) of the material is very small, radial junction allows for the fabrication of high-efficiency devices by decoupling the axis of light absorption and the axis of charge carrier separation. In a radial junction solar cell, light is absorbed along the main axis of the nanowire, while the charge carrier separation takes place in the radial direction which is just tens to a few hundreds of nanometers thick. In other words, in a radial junction solar cell, both charge carrier separation and light absorption can separately be optimized to achieve optimum performance of the nanowire solar cell.

In the last few years, there have been several reviews covering different aspects of nanowires and nanowire based solar cells [8-12, 14, 18-46]. Some of these reviews were more focused on the materials aspect of nanowire, whereas others were focused on optical or device behavior of nanowires. Even III-V nanowires solar cells have been widely reviewed[8, 10-12, 22, 33, 38-41, 44-46]. However, a dedicated review which simultaneously compares the optical and device behavior of axial and radial junction solar cells is still lacking. Because both axial and radial junction devices have their limitation as well as their complexities and advantages, it is essential to simultaneously compare both of them so that readers can get a better understanding on the choice of either an axial or radial junction solar cell.

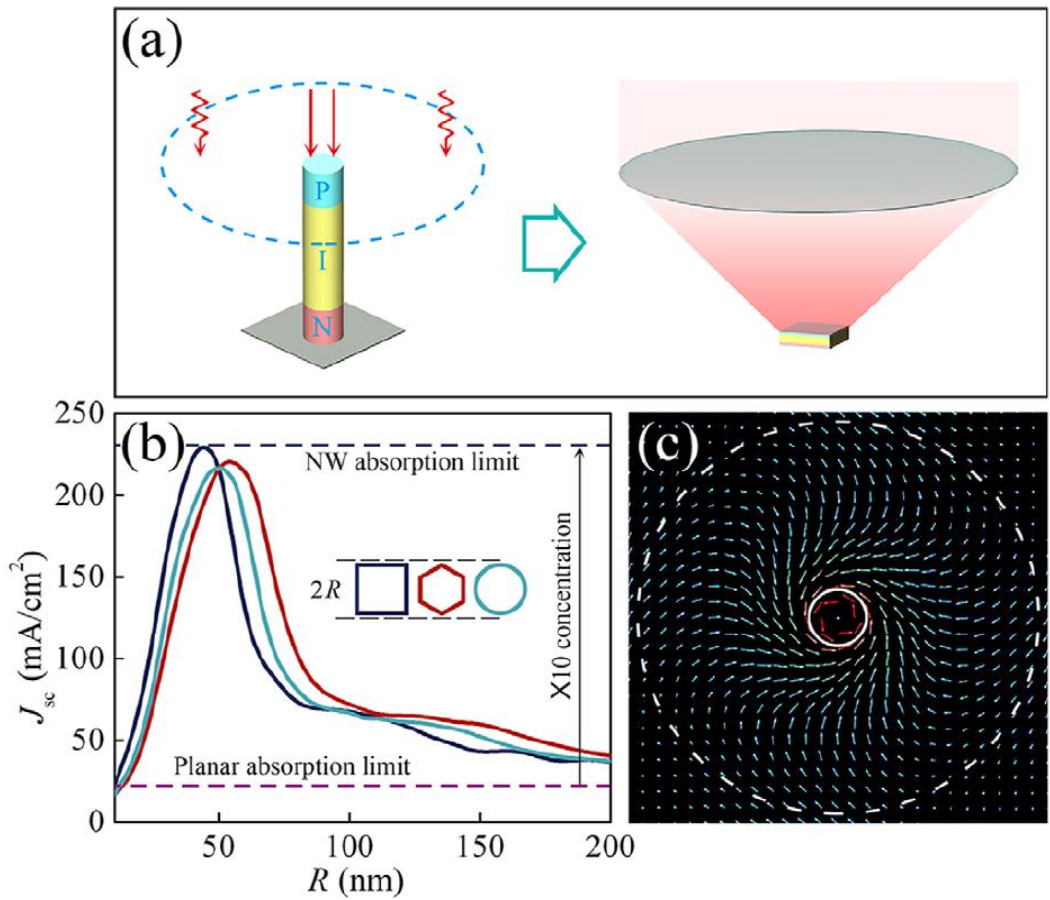

**Figure 2.** (a) Schematic showing a vertical single nanowire light concentration in the nanowire. (b) Plot showing that $J_{sc}$ can be substantially higher than the Shockley-Queisser limit for planar device due to light concentration in the nanowire. (c) Poynting vector at the middle plane normal to nanowire axis. The solid white circle is the outline of the nanowire cross section, while the white circle with broken line indicates the approximate position where the Poynting vectors start pointing towards the nanowire. *[Figures 2(a)-(c) have been reproduced from reference [52] under Creative Commons Attribution 3.0 International License.]*

---

In this review, we cover some of the recent results related to both axial and radial junction architectures and aim to compare the perfomance of these two types of devices. To keep the review more focused, we only discuss results of III-V solar cells; however, most of the concepts discussed are fundamental and can also be applied to other materials. We start by introducing the light absorption in single nanowires, followed by light absorption in nanowire arrays and methods to increase light absorption in both axial and radial junction solar cells. Next, we move on to discuss the device behavior of the solar cells, where we first introduce $V_{oc}$ and Sockley-Queisser limit of nanowire solar cells and then discuss the important concepts that are required to achieve high efficiency in both architectures. Finally, we discuss the opportunities and future prospects of nanowire solar cells, where we review some of the most crucial device concepts which are either complicated or not possible to realize in planar devices.

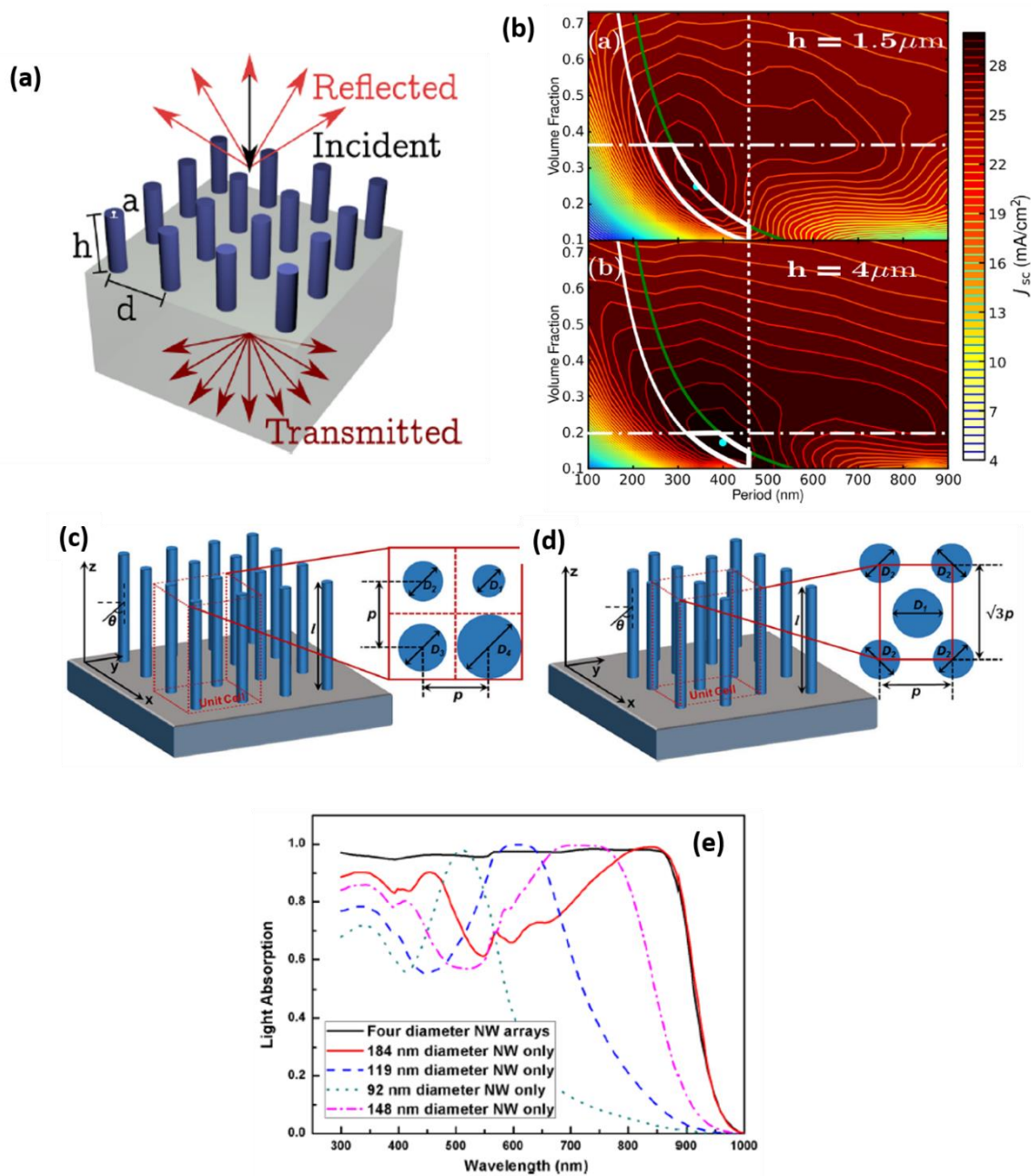

**Figure 3.** (a) 2-D schematic of a nanowire array with radius, a, pitch d, and height, h. (b) Optimization of $J_{sc(ideal)}$ for an InP nanowire array depicted in Figure 3 (a). (c) shows an InP nanowire array arranged in square lattice with nanowires of four different diameters. (d) shows an InP nanowire array arranged in hexagonal lattice with two different nanowire diameters. (e) shows that a combination of four different diameters arranged in a squre lattice shown in Figure 3 (c) is better than a square with nanowires of the same diameter. *[Figures 3(a) and 3(b) have been reprinted (adapted) with permission from ref [56]. Copyright (2014) American Chemical Society.]* *[Figures 3(c)-(e) have been reproduced from reference [58] under Creative Commons Attribution 4.0 International License.]*

---

**Light Management in Nanowires**

Nanowires are attracting a significant amount of attention because they can absorb much more photons per unit volume compared to their planar counterpart [9, 10, 12]. Before discussing absorption in a nanowire array, it is essential that we discuss the absorption in a single nanowire. It has been shown that for a single nanowire photon absorption cross-section can be significantly higher than their physical cross-section [42, 43, 47, 48]. The absorption enhancement in a single nanowire can be written as [39]:

$$AE\ (\%) = \frac{P_{abs}}{P_{inc} \cdot S} \times 100$$

where AE is the absorption enhancement, $P_{abs}$ is the power absorbed, $P_{inc}$ is incident power per unit surface area, and $S$ is the nanowire surface area. Often an AE exceeding 100% has been reported for single nanowires [37, 39, 42, 43, 49, 50]. As the absorption cross-section is higher than the physical cross-section, for single nanowires, the measured short circuit current density, $J_{sc}$ can exceed the Shockley-Queisser (SQ) limit [42, 51, 52]. Zeng et al. performed a detailed optoelectronic simulation of a verticalyl standing nanowire solar cell (see Figure 2(a)), and showed that the maximum obtainable $J_{sc}$ can exceed 200 mA/cm², which is several times higher than a planar solar cell (Figure 2(b)) [52]. Such a high absorption enhancement in single nanowires is a result of the "optical antenna effect". In other words, nanowires can efficiently couple electromagnetic radiation in free space into themselves, leading to excitation of guided optical resonant modes, which in turn leads to enhanced absorption [39, 42]. Figure 2(c) shows the Poynting vector plot for the single nanowire solar cell, which confirms that the nanowire leaky modes can efficiently couple light energy from several radii away from the nanowire. In another report, Krogstrup et al. fabricated a single GaAs nanowire solar cell with a $J_{sc}$ of 180 mA/cm$^2$, exceeding the Shockley-Queisser limit by a factor of six [42]. Furthermore, the absorption in the single GaAs nanowire (embedded in SU-8) was enhanced between 10 to 70 times compared to an equivalent thin film due to the larger effective absorption cross-section which was several times higher than the physical cross-section [42]. For a nanowire diameter of 380 nm, AE calculated using equation (1) is ~1200% [42]. They further confirmed the absorption cross-section enhancement by measuring a spatial photocurrent mapping of the single nanowire device at several wavelengths away from the nanowire.

*Absorption in Axial Junction Nanowire Arrays*

In a nanowire array, the optical modes become leaky and start interacting with each other to enhance absorption in a broader wavelength regime [44, 53-55]. Therefore, in a nanowire array, the nanowires should be arranged such that there is maximum overlap between the absorption cross-section of the individual nanowires of the array [44]. The absorption in a nanowire array can further be tuned by controlling the size, geometry, orientation, morphology, and the surrounding medium of individual nanowires. In this section, we review ways to tune the absorption in axial and radial junction nanowire arrays.

Absorption in a nanowire array depends strongly upon nanowire pitch (*P*) and radius (*R*), while it is only weakly dependant on the lattice arrangement of the array. Also, instead of discussing the effect of both pitch and radius separately, one can define a nanowire array in terms of its geometrical filling ratio (or volume fraction), which is given by the following equation:

$$Geometrical\ Fill\ Ratio\ (GFR) = \frac{\pi R^2}{P^2} \qquad (1)$$

Moreover, often researchers define absorption in a nanowire solar cell in terms of the maximum achievable short circuit current ($J_{sc(ideal)}$) for given solar cell structures. To calculate $J_{sc(ideal)}$, it is assumed that each photon absorbed creates one electron-hole pair.

To achieve maximum absorption in a nanowire array, the outward propagating mode needs to be minimized while simultaneously maximizing the resonant absorbing modes supported within the nanowires [56, 57]. Given that the optical effects in a nanowire are expected to be highly dispersive across the bandwidth of the solar spectrum, optimization of absorption in a nanowire array can often be very challenging [56]. To overcome the complex optimization barrier of absorption in a nanowire array, Sturmberg et al. [56] defined a nanowire parameter space, with pitch on the x-axis and GFR on the y-axis. Then they divided this parameter space into regions which either had a high reflection or a high transmission or did not support enough resonant absorption modes. In other words, to achieve maximum absorption in a nanowire array, the pitch and the GFR should lie outside these regions. This significantly reduced the optimization problem because the separate calculations of reflection, transmittance, and optical modes in a nanowire array (with given pitch and radius) were more straightforward compared to simultaneous calculations of all of the three parameters. Theoretical details on the calculation of transmission, reflection, and optical modes can be found in reference [56]. Figure 3(a) shows a 3-D schematic of a nanowire square array used by Sturmberg et al. to optimize the $J_{sc(ideal)}$ in an InP nanowire array. Figure 3(b) shows three different regions of the nanowire array parameter space for nanowires length of 1.5 μ and 4 μm, respectively. The region to the right of broken vertical dash line is a region of high transmittance, while the region above the broken dash-dot line is a region of high reflectance. Finally the regions outside the white and green lines are the regions that do not support enough resonant absorption modes. Therefore, the only region left after excluding all three unwanted regions is the region of maximum absorption, is shown by the thick solid white lines in Figure 3(b). Sturmberg et al. further reported that this optimization scheme is not dependent on the material of the nanowire array and can be applied to other materials as well [56].

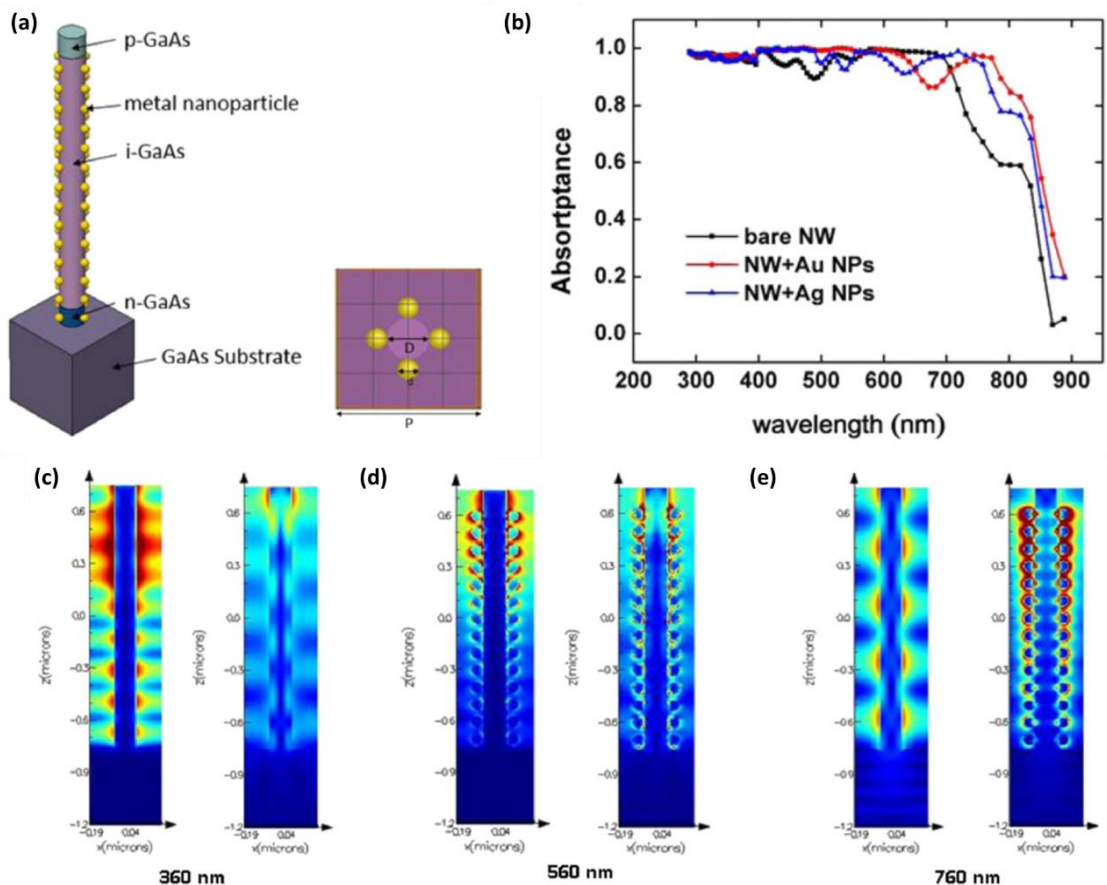

**Figure 4.** (a) A 3-D representation of p-i-n GaAs nanowire solar cell decorated with metal nanoparticles. (b) Comparative absorption vs. wavelength spectra for bare GaAs nanowire and GaAs nanowire coated with gold and silver nanoparticles. (c)-(e) show light concentration in nanowire with (right) and without (left) metal nanoparticles at different wavelengths. *[Figures 4(a) –(e) have been reprinted (adapted) from reference [67] under Creative Commons Attribution 4.0 International License]*

Though nanowire arrays show higher absorption compared to their planar counterpart in volumetric terms, yet, there are several different ways in which absorption in the arrays can be further improved. As mentioned earlier for absorption in single nanowires, the optical resonant modes are highly dependent on nanowire diameter and therefore, a nanowire array consisting of single nanowire diameter cannot achieve optimal absorption over a broadband wavelength range. Wu et al. [58, 59] in two separate contributions have shown that the absorption in a nanowire array can be maximized by the utilization of an array lattice with nanowires of different radii. Figure 3(c) and 3(d) shows two such lattice arrangements with nanowires of different radii. While the square lattice consists of nanowires of four different diameters, the hexagonal lattice consists of two different diameters. Figure 3(e) shows that a square lattice with four different radii of nanowires shows broader absorption compared to a square lattice with nanowires of one diameter. Another way to achieve broadband absorption is through the tapering of the nanowires (nanocone). In a tapered nanowire, the diameter of the nanowire varies along the length of the nanowire; thereby, several wavelength-specific resonant modes can be excited at the same time to achieve the broadband absorption [57, 60-62].

Another approach to enhance absorption in a nanowire array is through the utilization of plasmonic metal nanoparticles. In the metal nanoparticles, localized surface plasmon resonant modes can be excited to concentrate and transfer incident light energy into the nanowire. The use of plasmonic nanoparticles to enhance absorption in oxide nanostructure is well-known [63-66]. In comparison, plasmon enhanced absorption in III-V nanostructures is less studied. Recently, Li et al. have shown that the absorption in nanowire arrays can be enhanced in the presence of metal nanoparticles [67]. Figure 4(a) shows an axial p-i-n GaAs nanowire decorated with metal nanoparticles. Figure 4(b) shows the comparative absorption spectra of bare GaAs nanowire and nanowire decorated with gold and silver nanoparticles. It is evident that in the presence of nanoparticles, nanowire absorption is enhanced in the longer wavelength regime, compared to bare GaAs nanowire. They also showed that the size of metal nanoparticles can affect the absorption of the nanowire, and needs to be efficiently tuned to achieve local field enhancement while also minimizing reflection from the metal nanoparticles. Figures 4(c) – 4(e) show the normalized electric field distribution in the nanowire with and without nanoparticles at different wavelengths. The concentration of energy (proportional to $|E|^2$) around the nanowire is higher in the presence of metal nanoparticles at 760 nm wavelength, which is also reflected in enhanced absorption at 760 nm wavelength when the GaAs nanowire is coated with metal nanoparticles. Review of plasmonic enhancement in nanowires using metal nanoparticles can be found in reference [31].

Additionally, researchers have also shown that significant improvement in absorption of a semiconductor can be realized in the presence of a non-absorbing dielectric shell [15, 57, 68-70]. This enhanced absorption in the presence of a non-absorbing shell is a result of increased scattering cross-section, reduced electric field screening, and the optical antenna effect [57, 69, 70]. The basic idea behind the use of a dielectric shell is to reduce the dielectric function contrast between the nanowire core and the air such that the screening of incident electric field can be reduced, while also maximizing the supported absorption modes in the core. To achieve optimum absorption in the nanowire core, the thickness of the dielectric layer should be optimized such that it maximizes the supported absorption modes in the nanowire core while also increasing the scattering cross-section. There have been reports on improvement in absorption of different semiconductor nanowire arrays such as Si [70], InAs [69], GaAs [71], InP [15, 68, 70], etc. For example, Anttu et al. performed a detailed theoretical and experimental investigation on InAs-$Al_2O_3$ core-shell nanowires and showed that $Al_2O_3$ can significantly improve the reflectance from InAs nanowires for an optimized $Al_2O_3$ thickness [69].

*Absorption in Radial Junction Nanowire Arrays*

In the case of a core-shell nanowire, the absorption follows an almost similar trend as an axial nanowire if both core and shell are made of the same material. However, there are often cases where the core and the shell are formed of different materials. In such cases, both core and shell need to be simultaneously optimized to achieve optimum optical absorption, without compromising device performance. Further, the presence of a shell over an absorbing core can significantly change the absorption in the core due to mismatch in refractive indices and absorption coefficients between the core and shell. For example, Figure 5(a) shows a 2-D schematic of a nanowire array solar cell with GaAs nanowires embedded in P3HT hole selective contact, and the p-n junction is formed radially [72]. Figure 5(b) shows that even for a filling ratio as low as 0.05, GaAs/air structure can achieve high absorption in the longer wavelength regime of 450-800 nm. However, after P3HT deposition (see Figure 5(c)), the absorption in longer wavelength region is reduced, while the absorption in the shorter wavelength region is improved. The improved absorption in shorter wavelength region is a result of high absorption in P3HT, which has an absorption edge at 630 nm. It means that P3HT is absorbing a large portion of lower wavelength photons before they reach GaAs. However the photons absorbed by P3HT are lost through recombination, and hence Wu et al. reported that to achieve 22 mA.cm$^{-2}$ photocurrent, the required GaAs filling ratio was ~0.7 [72]. This is in contrast to other reports on GaAs solar cells which showed that GaAs-air nanowire arrays can achieve more than 22 mA/cm$^2$ for filling ratio less than 0.2 [58, 73, 74].

In another report, Huang et al. showed a similar decrease in absorption in GaAs/Al$_{0.8}$Ga$_{0.2}$As core-shell solar cell arrays (see Figures 5(d) and 5(e)) [75]. They showed through simulation that after the deposition of Al$_{0.8}$Ga$_{0.2}$As shell, there was a decrease in absorption in the GaAs core, and J$_{sc(ideal)}$ was reduced from 25.93 mA/cm$^2$ to 23.13 mA/cm$^2$, for an array filling ratio of 0.196. However, they also showed that the presence of Al$_{0.8}$Ga$_{0.2}$As passivation layer was critical in achieving high efficiency in GaAs nanowire solar cells, from device point of view.

Nevertheless, a reduction in the absorption of a core after deposition of a shell is not always the case. In fact, we have recently shown that in an AZO/ZnO/p-InP heterojunction solar cell, an optimized thickness of oxide shell can significantly improve the absorption in InP nanowire, while also improving its device characteristics [15]. Figure 6(a) shows a 2-D schematic of the proposed device structure with p-type InP as core and AZO/ZnO as a shell. We show that for 90 nm of shell thickness (see Figure 6(b)), the $J_{sc(ideal)}$ reaches to more than 32.4 mA/cm$^2$ for filling ratio as low as 0.09. In comparison, the $J_{sc(ideal)}$ for bare InP remains at 29.4 mA/cm$^2$ for a similar filling ratio. The increase in $J_{sc(ideal)}$ is a direct result of improved absorption (see Figure 6(c)) in the shorter wavelength regime due to improved electric field confinement in InP nanowires in the presence of an oxide shell. Most importantly, AZO/ZnO also improves the electrical characteristics of the solar cells. The simulation results have further been verified by the fabrication of proposed radial junction solar cell in reference [76]. The improved optical behavior stems from the higher confinement of electric field in the presence of an oxide shell, whereas an improved electronic behavior is a combined result of passivation, electron selectivity, and a large built-in electric field, as discussed later.

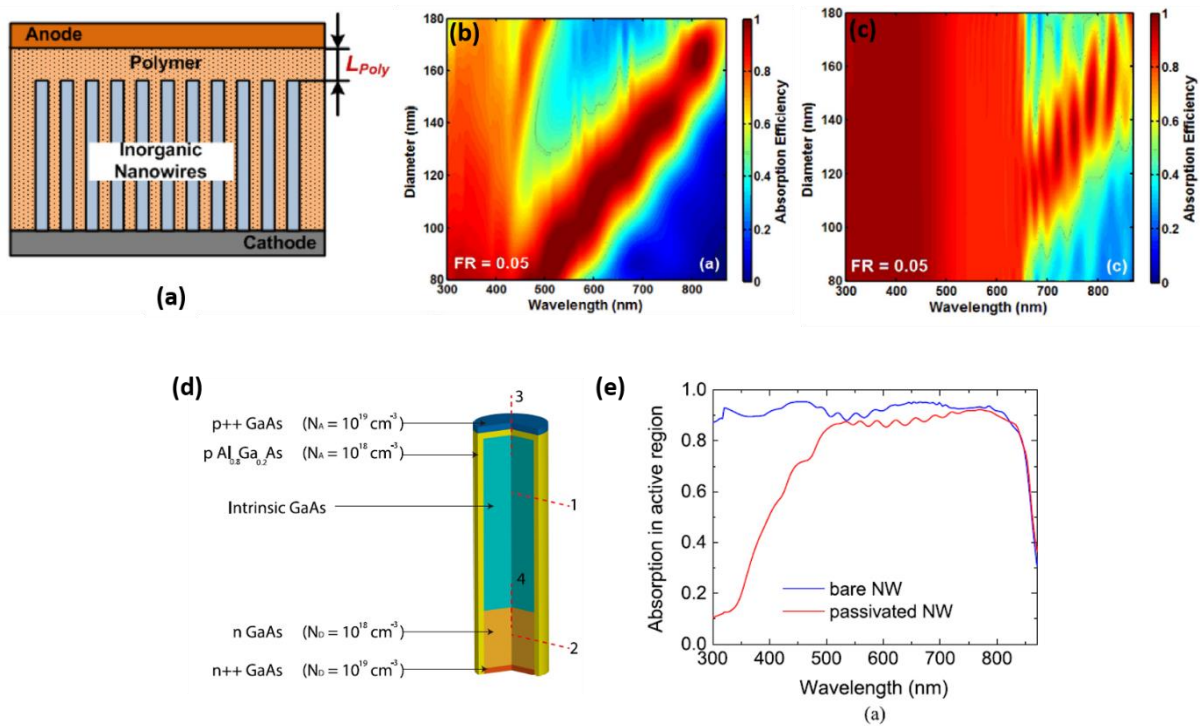

**Figure 5.** (a) 2-D schematic of a GaAs nanowire array embedded in P3HT. (b) and (c) show the absorption vs wavelength for different nanowire radii when the array is in air and embedded in

P3HT, respectively. (d) Schematic of a GaAs nanowire with an AlGaAs shell. (e) shows that the absorption reduces below 500 nm after deposition of the AlGaAs shell. *[Figures 5(a)-(c) have been reprinted (adapted) with permission from ref [72] Copyright (2019) OSA.] [Figures 5(d)-(e) have been reprinted (adapted) with permission from ref [75]. Copyright (2019) IEEE.]*

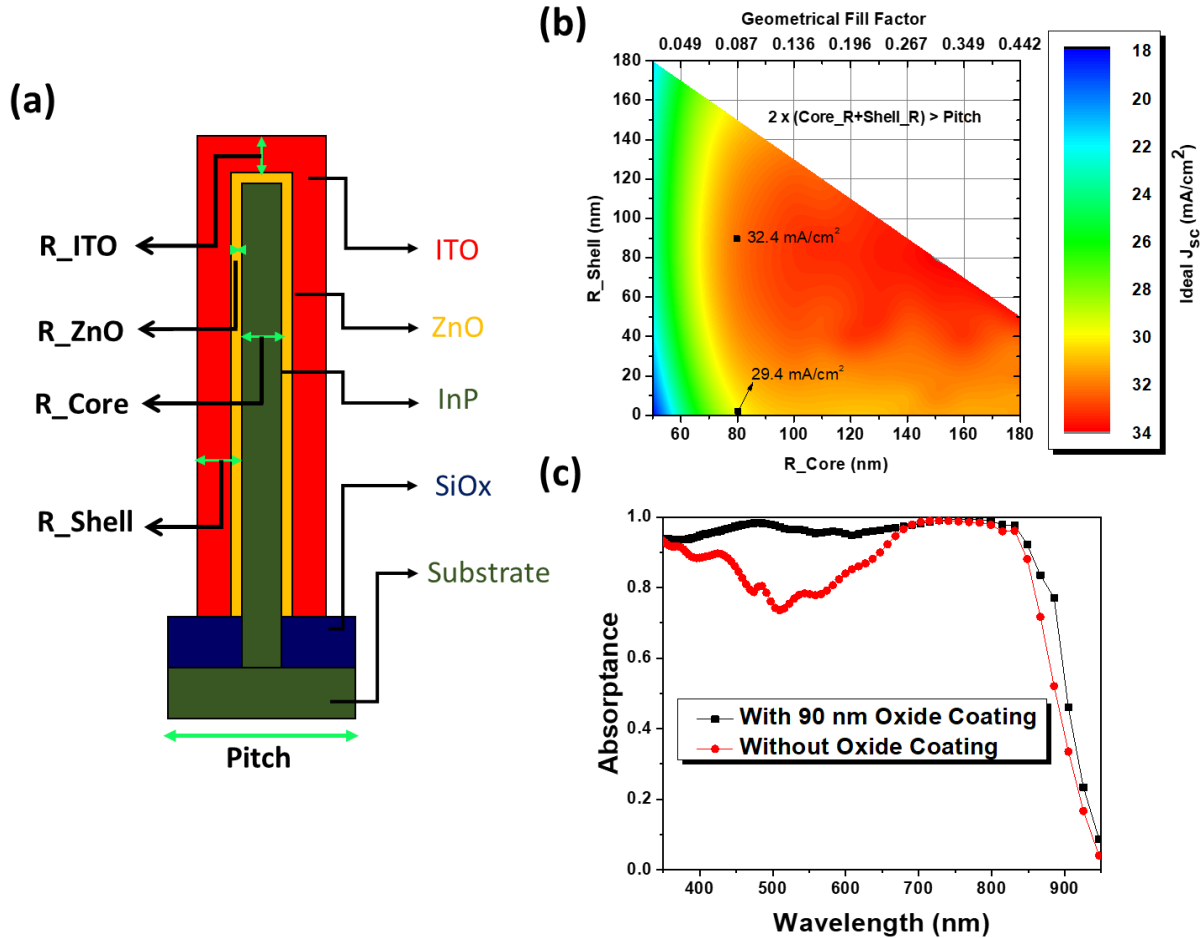

**Figure 6.** (a) 2-D schematic of a radial junction nanowire solar cell using p-InP as core and AZO/ZnO as shell. (b) Calculated $J_{sc(ideal)}$ as a function of shell thickness (R_Shell) and nanowire core radius (R_Core). (c) Absoprtion vs wavelength plot for InP nanowire with and without an oxide shell. *[Figures 6(a)-(c) have been reprinted (adapted) with permission from ref [15] Copyright (2019) IEEE.]*

**Device design**

*$V_{oc}$ of nanowire solar cells:*

In 1961, Shockley and Queisser (SQ) provided a theoretical framework to predict the maximum achievable performance of a planar solar cell under the principle of detailed balance [77]. Detailed balance limit mainly concerns with the absorption and emission of light in a given semiconductor. However, for semiconductor nanowires, both their absorption and emission can be tuned by their dimensions and shape, and therefore, there has been an increased research interest in understanding the upper limit of efficiency of nanowires [51, 78, 79]. Yunlu et al. have shown that the improvement in the efficiency of nanowire solar cells under SQ-limit is strictly due to improvement in $V_{oc}$ [79]. This improvement in comparison to the planar device is a result of light concentration by the nanowires, as discussed below.

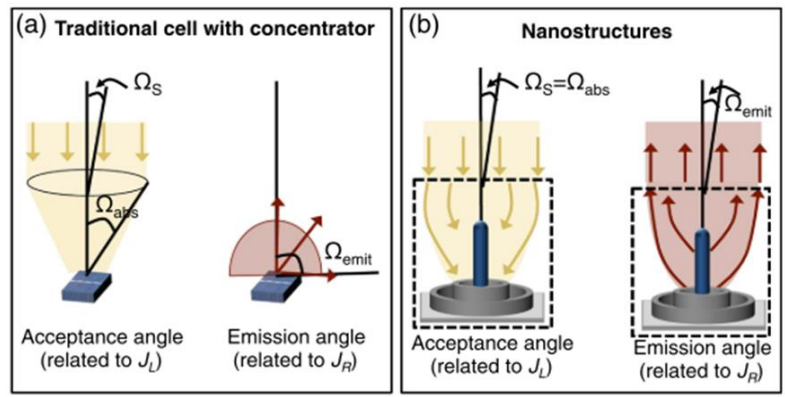

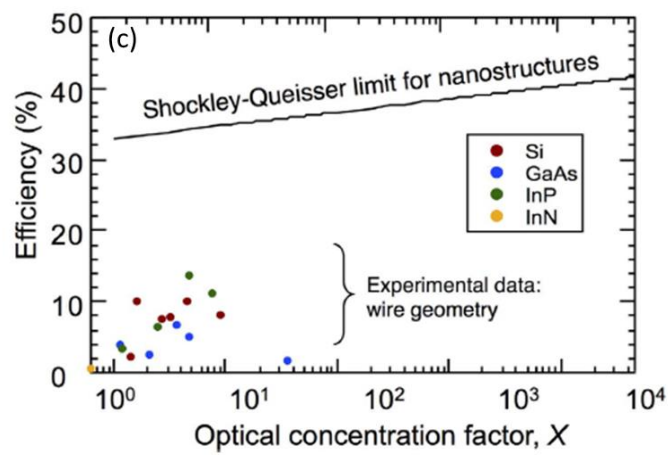

**Figure 7.** (a) and (b) compares the concentration effect in planar and nanowire solar cells. In a traditional planar solar cell, an optical concentrator and a back reflector are used to match the etendue of incident and emitted photons. A nanowire solar cell with back metal reflector has the same effect, because nanowires have the inherent ability to concentrate light. *[Figures 7(a)-(c) have been reproduced from reference [79] under Creative Commons Attribution 4.0 International License.]*

---

In Boltzmann's approximation, the open-circuit voltage ($V_{oc}$) of a solar cell operating at its maximum efficiency is given by:

$$q \cdot V_{oc} = E_g \left(1 - \frac{T_{cell}}{T_{sun}}\right) + k_b T_{cell} \cdot \ln\left(\frac{\gamma_{sun}}{\gamma_{cell}}\right) - k_b T_{cell} \left(\frac{\Omega_{emission}}{\Omega_{incident}}\right)$$

$$\gamma_{sun} = \frac{2kT_{sun}}{c^2 h^3} \left(E_g^2 + 2kT_{sun}E_g + 2k^2 T_{sun}^2\right)$$

$$\gamma_{cell} = \frac{2kT_{cell}}{c^2 h^3} \left(E_g^2 + 2kT_{cell}E_g + 2k^2 T_{cell}^2\right)$$

where, $E_g$ is the material bandgap, $T_{cell}$=300 K, $k_b$ is the Boltzmann's constant, $T_{sun} = 5700\ K$ is the temperature at the sun's surface, $\Omega_{emission}$ is the étendue of the emitted photon, and $\Omega_{incident}$ is the étendue of the incident photon. The first term in the above equation corresponds to the Carnot efficiency of converting a photon with energy $E_g = \hbar\omega$ (where $\omega$ is the angular frequency) to electrostatic energy. The second term in the above equation is associated with the irreversible entropy generated due to photon cooling. This occurs due to the temperature difference between the absorbed and the emitted photon. Finally, the third term is a result of entropy generation due to the difference between the étendue of the incident photon to the étendue of the emitted photon. The third term represents a voltage loss of ~280 mV and can be eliminated if $\Omega_{emission}$ becomes equal to $\Omega_{incident}$. Modification of the directionality of absorption and emission in a solar cell is a well-known method for improving the $V_{oc}$ of a device [80-83]. In a planar solar cell, the recovery of entropy due to a mismatch between absorption and emission angle is achieved through the use of a metal back reflector and optical concentration [80, 82]. Utilization of a metal reflector reduces the emission angle from $4\pi$ to $2\pi$, whereas utilization of concentrator optics is required to exceed the emission angle from the sun's solid angle to match the emission angle. An adequately designed nanowire solar cell with a metal reflector can have the same effect because of in-built optical concentration. Figures 7(a) and 7(b) shows a schematic of the comparison between the concentration of light in a planar solar cell using optics and self concentration of light in nanowires. Figure 7(c) shows the enhancement in efficiency for different levels of optical concentration. Colored dots on Figure 7(c) show the experimental results of nanowire solar cells with different levels of in-built light concentrations. For further details, readers are referred to reference [79].

Although the Shockley-Queisser limit predicts a higher $V_{oc}$ and therefore higher efficiency in nanowire solar cells compared to their planar counterpart, yet the reported $V_{oc}$ for nanowire solar cells remains significantly below the predicted value. This is mainly because the performance of nanowire solar cells depends significantly on the device geometry in 3-D along with p-n junction position, design, and depth, which complicates the optimization of these devices. Another problem is the large surface area of a nanowire, which exposes it to high surface recombination, and passivation of nanowires becomes inevitable to achieve high performance. In the next section, we discuss some of these metrics on which the device performance depends on.

*Device design of axial junction nanowire solar cells*

In an axial junction solar cell, the position and depth of the junction can have a significant effect on charge carrier separation and collection. In the case of nanowires, it becomes extremely important to achieve a p-n junction near the top of the nanowire because most of the carriers are generated in the top segment of nanowire, and if the junction is very far from top segment, most of the carriers will recombine and will not constitute any current due to the low diffusion length

[84]. In a detailed study of p-i-n junction nanowire solar cells, Gao et al. have shown the importance of junction position on the efficiency of an axial junction solar cell [84]. They show that by changing the doping profile in nanowires, the position and depth of p-n junction can be tuned such that the junction forms near the top of the nanowire [84]. They studied three different doping combinations: (I) $p^+$-$n^-$-n, (II) $p^+$-p-$n^-$-n and (III) $p^+$-p-$n^-$-n and found that only for sample (III), the junction was formed near the top of the nanowire, whereas for other two the junction formed away from the top of the nanowire. In almost similar work on GaAs nanowire solar cells, Otnes et al. reported that they were able to improve the $J_{sc}$ of solar cell from 5 mA/cm$^2$ to 25 mA/cm$^2$ just by changing the position of the junction to the top of the nanowire [85].

Another essential parameter required to achieve high efficiency in a p-i-n nanowire solar cell is the control of the length of the doped regions. Wallentin et al. in their highly celebrated paper showed that the control of the length of the top-most n-type segment was of significant importance to achieve high efficiency in the nanowire solar cell [3]. Figure 8(a) shows the 2-D schematic of a nanowire solar cell on which the measurements were performed to ascertain the importance of the length of the top n-type segment. Figures 8(b) and 8(c) respectively show the effect of the length of top n-segment on $J_{sc}$ and quantum efficiency. It is quite evident that for four samples that were investigated, the best efficiency was achieved when the length of the top n-type segment was 60 nm. Any further increase or decrease in the length of n-type segment led to a substantial deterioration in the $J_{sc}$. Also, from quantum efficiency measurements, it can be inferred that the external quantum efficiency (EQE) in the shorter wavelength regime was severely affected by the duration of the growth of the top n-type segment. This is due to shorter wavelength light being mainly absorbed at the top segment of the nanowire. Furthermore, from our knowledge of planar junction solar cell, the top n-type segment should be as thin as possible to reduce parasitic absorption of light by the n-type segment itself. In other words, because charge carriers that are generated in a heavily doped top n-type segment are lost through non-radiative recombination, the length of n-type segment should be as thin as possible. However, in the case of nanowire solar cells, this problem is more severe compared to their planar counterpart because the absorption in top few nanometers of the nanowires is significantly higher compared to a planar structures, and therefore leads to a decrease in photogenerated current.

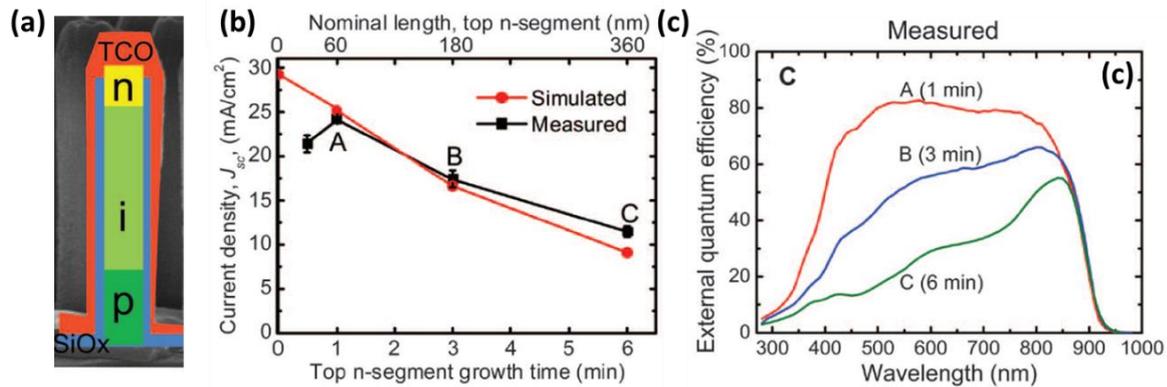

**Figure 8.** (a) Schematic of an axial junction nanowire solar cell used for studying the effect of top n-type segment on $J_{sc}$ of the device. (b) and (c) show that effect of n-type top segment thickness and growth time on $J_{sc}$ and external quantum efficiency, respectively. *[Figures 8(a)-(c) have been reproduced from reference [3]. Copyright (2013) American Association for the Advancement of Science (AAAS). ]*

Similar to planar solar cells, nanowire solar cells may also be limited by contact recombination. To show the importance of contact recombination in nanowire solar cells, Chen et al. performed a series of simulations and showed that when a p-i-n or p-n junction nanowire solar cell is in direct contact with the metal, it can lose a large portion of photogenerated current through recombination at the front contact[86]. To overcome contact recombination losses, they employed GaP layer as a carrier selective contact. In recent times, carrier selective contacts have emerged as an efficient way to reduce contact recombination [87]. A carrier selective contact acts as a semi-permeable membrane which allows only one kind of carrier (e.g., electrons) to pass through while blocking the other kind of carrier (e.g., holes). They reduce the contact recombination by impeding the flow of minority carrier to the contacts. Chen et al. showed that through the utilization of a wide bandgap GaP, the contact recombination in both p-i-n and p-n junction InP nanowire solar cells can be reduced to achieve high efficiency [86]. They also reported that just by using GaP on the top segment of p-i-n nanowires, there was an improvement of ~3 mA/cm$^2$ in $J_{sc}$; however, the $V_{oc}$ remained almost unchanged. When both the n-GaP and p-GaP were respectively applied on the top and bottom of the p-i-n nanowires, both the $J_{sc}$ and $V_{oc}$ improved significantly compared to nanowires without GaP [86]. Another important observation that they made was that the effect of carrier selective contact was more pronounced on p-n junction as compared to p-i-n junction, especially when bulk minority carrier lifetime of the nanowires was low. They postulated that this effect was due to minority carriers generated in the intrinsic region of the nanowire having to travel a longer distance, compared to carriers generated in either the n-type or p-type segment of the nanowire [86]. Other ways to tackle the problem of contact recombination is through the implementation of a back or front surface field. Aberg et al. have shown that the implementation of a back surface field (BSF) along with surface passivation is critical in achieving high efficiency in axial GaAs nanowire solar cells [88]. Other references on the importance of contact recombination in the context of a nanowire solar cell can be found in references [33, 73, 89-91].

Other than device design, one of the most critical parameters to achieve high efficiency in a nanowire solar cell is surface passivation. In the case of nanowires, passivation becomes almost inevitable because of the large surface-area-to-volume ratio. Therefore, there has been considerable interest in studying the effect of passivation on nanowires [3, 22, 68, 87, 92-101]. Most recently, Otnes et al. have shown that even for the most optimized nanowire solar cell in terms of device design, passivation is indispensable to reach its full potential [85]. They started their experiment by optimization of growth conditions, junction depth, and position, and in the final step, they introduced the passivation layer. For passivation, they utilized silicon dioxide grown using two different precursors and named the samples "oxide 1" and "oxide 2". Figure 9(b) shows the effect of "oxide 1" and "oxide 2" on the light IV characteristics of axial junction InP nanowire solar cells. For deposition of "oxide 1", they used bis(diethylamino)silane as the precursor, whereas, to deposit "oxide 2" they used trimethylaluminum (TMAl)/tris(tert-butoxy)silanol (TTBS) chemistry [85]. Just by optimization of the oxide deposition conditions, they were able to reduce the dark saturation current ($J_0$) by two orders of magnitude (see Figure 9 (a)). As a result of improved passivation and reduced dark current, both the $V_{oc}$ and the $J_{sc}$ were improved to achieve an efficiency of ~15%. Furthermore, they found that "oxide 1" was less suitable for passivation of InP compared to "oxide 2" and claimed that this is due to "oxide 2" having less Al contamination [85]. In another report, Cui et al. stressed the significance of nanowire cleaning [101]. They claimed that just by optimization of the nanowire cleaning procedure, they were able to reduce the dark current by four orders of magnitude and increase the efficiency of the solar cell from a fraction of a percent to 11%. Figure 9 (c) shows the best IV corresponding to the best device fabricated using the piranha cleaning procedure, whereas Figure 9 (d) shows that dark current was reduced by several orders of magnitude after treatment with the piranha solution for 30s [101].

*Device design of radial junction nanowire solar cells*

In comparison to axial junction nanowire solar cells, studies on radial junction nanowires are relatively sparse, mainly due to the complexity involved in the growth, doping, and fabruication of the core-shell nanowire structures. Li et al. covered some of the essential aspects of silicon based radial junction solar cell in reference [102]. Other important works on core-shell and radial junction nanowire solar cells can be found in references [103-105]. In the next section, we will try to cover some of the most important aspects required for achieving high efficiency in radial junction nanowire solar cells.

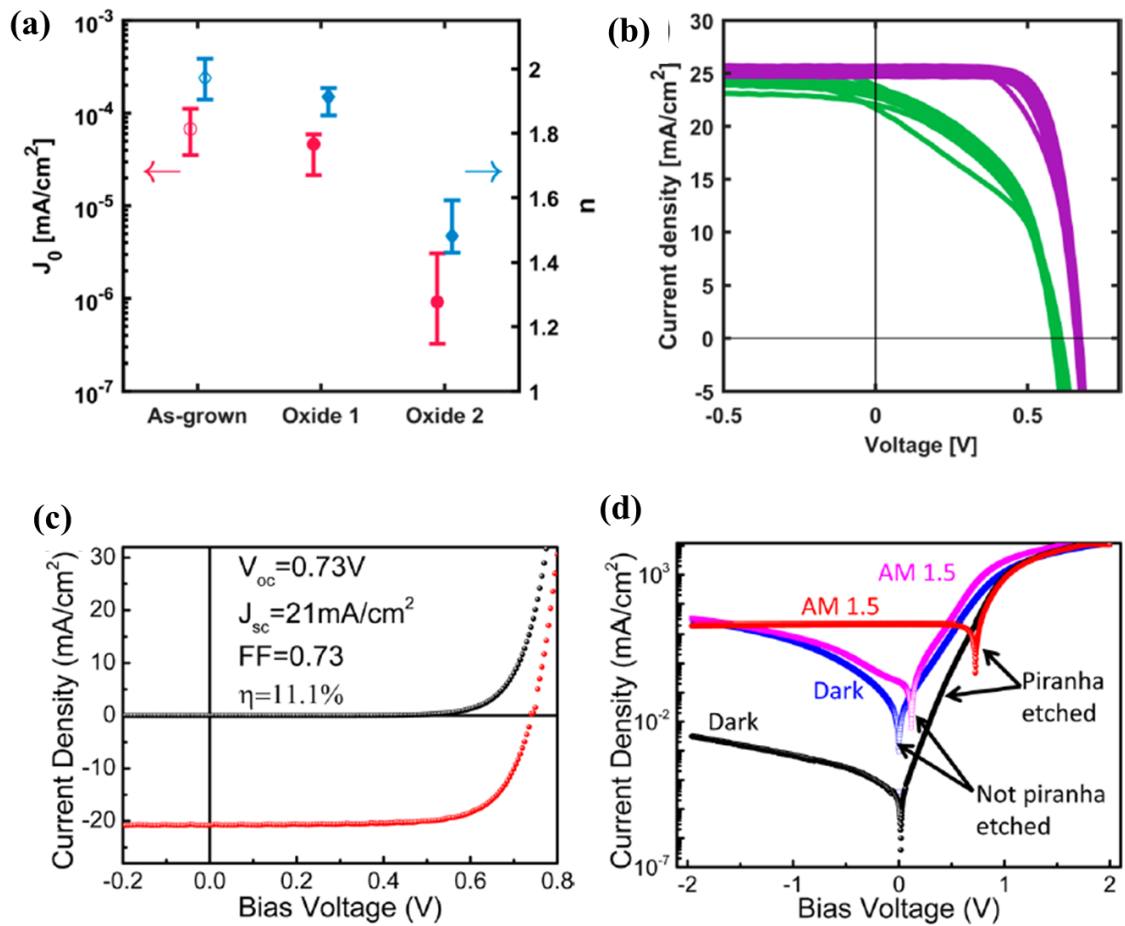

**Figure 9.** (a) Dark saturation current density for as-grown and oxide passivated GaAs solar cells. Oxide1 and Oxide2 represent $SiO_2$ deposited using different precursors. (b) IV curve of GaAs solar cell with Oxide1 (green) and Oxide2 (magenta) passivation. (c) Dark and light IV curves of best device obtained thorough a cleaning procedure using piranha etching. (d) Dark IV vs. voltage curve for nanowire solar cells with and without piranha etching. *[Figures 9(a) and (b) have been reprinted (adapted) with permission from ref [85]. Copyright (2018) American Chemical Society.] [Figures 9(c) and (d) have been reprinted (adapted) with permission from ref [101]. Copyright (2013) American Chemical Society.]*

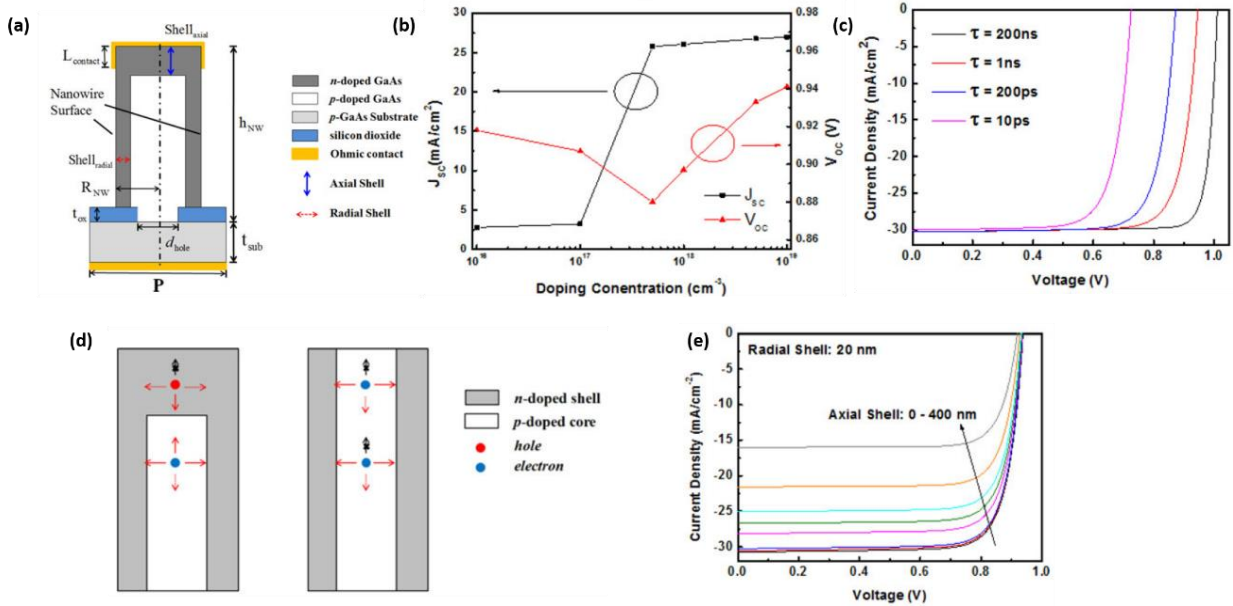

**Figure 10.** (a) 2-D schematic of a core-shell GaAs nanowire with the shell forming both radially and axially. (b) Effect of core doping concentration on $J_{sc}$ of the core-shell solar cell depicted in Figure (a). (c) Effect of minority carrier lifetime on IV of the core-shell solar cell depicted in Figure (a). (d) The charge carrier transport in a solar cell shown in Figure 10(a) can either be purely radial (right) or a combination of axial and radial (left). (e) Effect of axial shell thickness for a fixed radial shell thickness of 20 nm. *[Figures 10(a)-(e) have been reprinted (adapted) with permission from ref [16] Copyright (2015) IEEE.]*

Li et al. performed a detailed simulation of a GaAs based core-shell homojunction. They showed that the overall performance of the solar cell depends heavily on doping in the core of the nanowire [16]. Figure 10(a) shows a 2-D schematic of core-shell nanowire solar cell with the shell forming both axially and radially. Formation of both axial and radial shell is very normal in case of core-shell nanowires because of the low selectivity in deposition on the sidewalls and the top of the nanowire [16]. Figure 10(b) shows the effect of core doping on $J_{sc}$ and $V_{oc}$. It is quite evident that the $J_{sc}$ of the solar cell is extremely low for low or moderate doping of the nanowire core. The authors argued that when the doping in the core is low or moderate, the junction is formed between the substrate and the core, rather than between the core and the shell [16]. Additionally, the core is fully depleted of the charge carriers as the depletion region extends into the core. In other words, when core doping is low, the solar cell behaves like an axial junction solar cell (with the junction forming between the substrate and the core) with a fully depleted core. Therefore, to achieve a sufficiently high $J_{sc}$, both core and shell should be heavily doped so that the core is not depleted of charge carriers and maintains a sufficiently high radial built-in electric field for charge carrier separation [16]. Another critical parameter that affects a solar cell is its bulk minority carrier lifetime. The dependence of a solar cell performance on minority carrier lifetime is very well known, yet for the core-shell structure it is much more interesting to study because the carriers need to travel the smaller distance before being collected by the contacts. As expected, the proposed core-shell solar cell was able to achieve ~13% efficiency even when the minority carrier lifetime is as low as 10 ps [16]. Moreover, they found that the biggest problem in a core-shell nanowire solar cell was not its bulk minority carrier lifetime but the surface recombination velocity (SRV) at the core-shell interface. They reported that the efficiency was reduced significantly when the surface recombination velocity at core-shell interface exceeded $10^3$ cm/s. Another critical parameter which defined the overall performance of a radial junction nanowire solar cell is the thickness of the axial or radial shell [16]. Figure 10(d) shows that when both axial and radial junctions are formed, there is competition for charge carriers to separate either axially or radially, depending on where the charge carriers have been generated. They find that within the range of the thicknesses investigated, the axial shell can have a huge effect on the overall efficiency [16]. During the simulation, they varied the thickness of the radial shell from 0-20 nm, while that of the axial junction from 0-400 nm, and found that the overall efficiency was reduced significantly with increasing axial shell thickness, especially when the minority carrier lifetime of the core was low [16]. Figure 10(e) shows the effect of axial shell thickness on device performance when the radial shell was fixed at 20 nm.

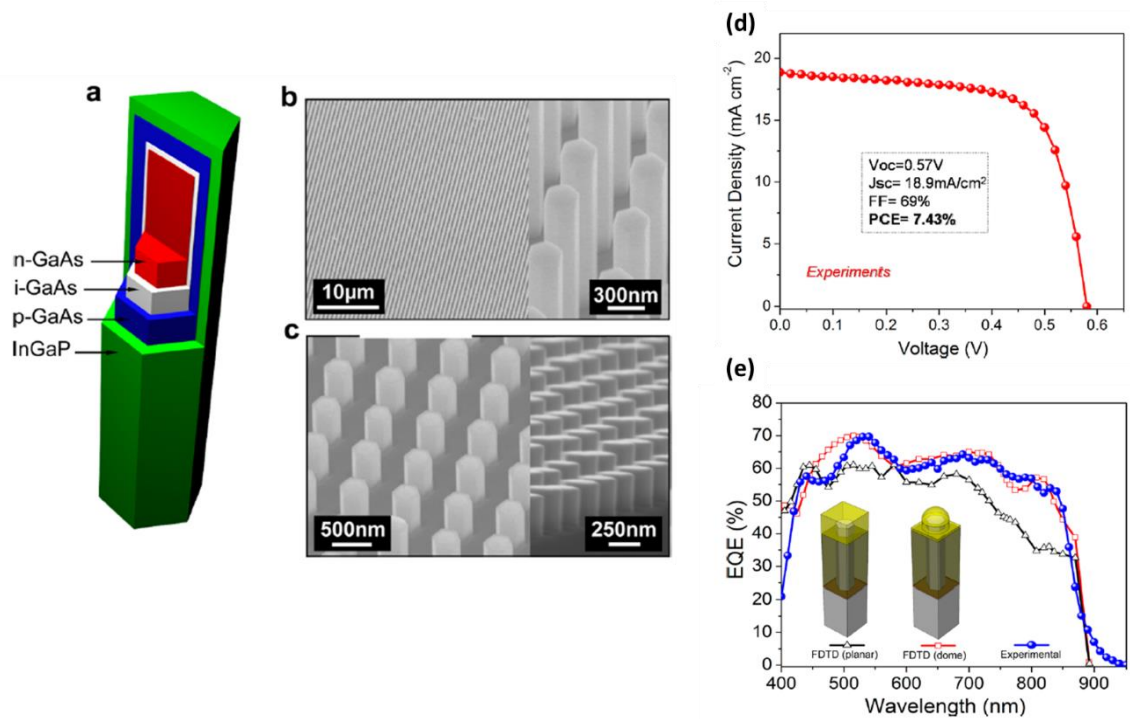

**Figure 11.** (a) Schematic of a homojunction GaAs core-shell nanowire solar cell with InGaP passivation. (b) and (c) shows the SEM images of the solar cell depicted in 11(a). (d) IV curve of the best core-shell solar cell. (e) Comparative EQE of nanowire solar cells with planar ITO and dome shaped ITO. *[Figure 11 has been reprinted (adapted) with permission from ref [106]. Copyright (2013) American Chemical Society.]*

---

Although there have been some attempts to realize high-efficiency III-V radial junction nanowire solar cells, most of the efforts have been in vain with efficiencies remaining significantly lower than their axial junction counterpart [45]. Though detailed experimental studies on the reasons for this discrepancy are limited, it has been postulated that it may be because of the requirement of achieving high core and shell doping while simultaneously minimizing the surface recombination velocity at the interface [15, 45]. Also, in general, III-V core-shell solar cells are grown using MOCVD and the shell is grown on top of the core at a relatively high temperature which may lead to cross-contamination and out-diffusion of dopants, leading to high recombination and reduced efficiency. Nonetheless, Mariani et al. reported a core-shell solar cell with a p-i-n homojunction and an InGaP passivation layer (see Figure 11(a)-(c)) [106] and claimed to achieve a surface state density as low as $N_t = 1\times10^{10}$ cm$^{-2}$eV$^{-1}$, while achieving an efficiency of 7.43% with a $V_{oc}$ and $J_{sc}$ of 0.57 V and 18.9 mA/cm$^2$, respectively (see Figure 11 (d)). They also showed that the efficiency of the solar cell improved by using a dome-shaped ITO compared to planar ITO because of enhanced absorption (Figure 11(e)). A similar effect of dome-shaped ITO has also been reported for axial junction solar cells [80].

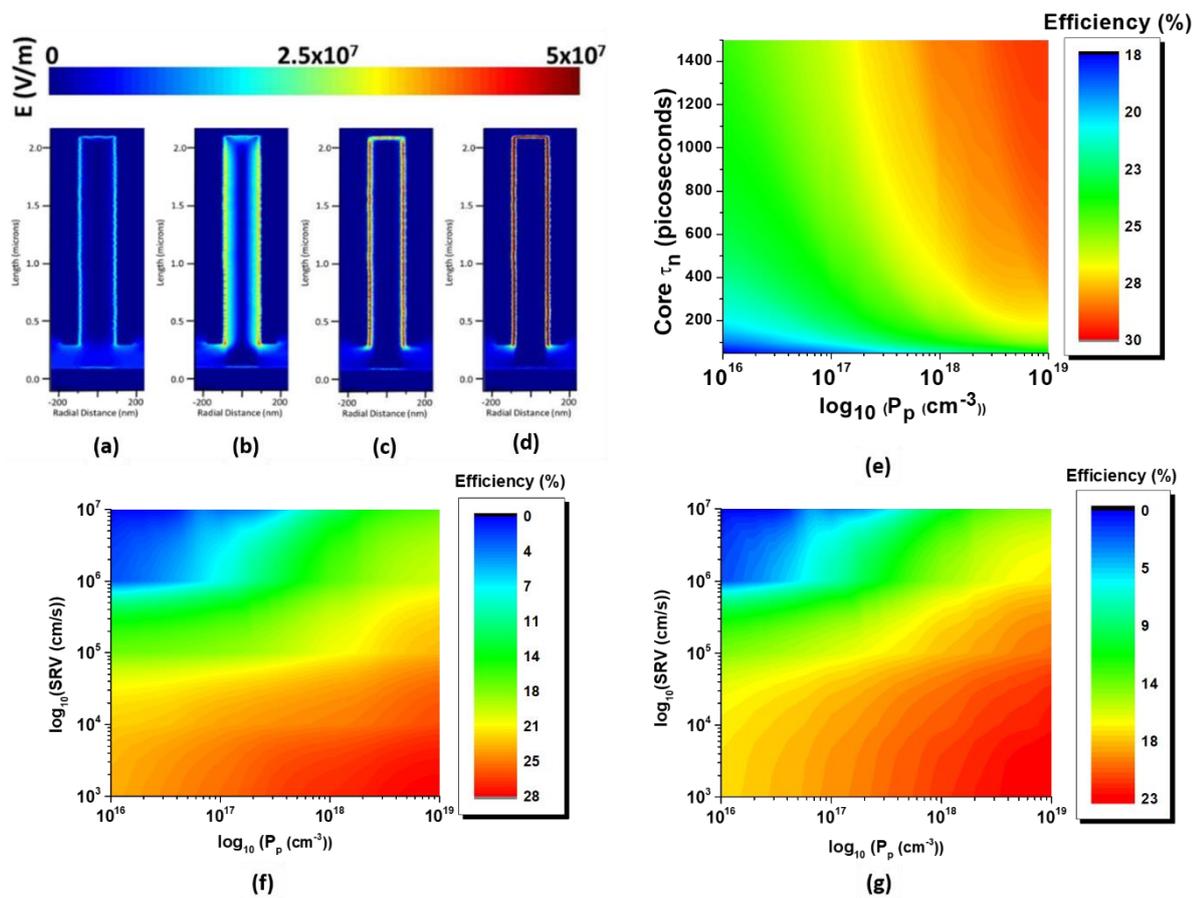

**Figure 12.** Simulated built-in electric field in radial heterojunction for core doping of (a) $1\times10^{16}$ cm$^{-3}$, (b) $1\times10^{17}$ cm$^{-3}$, (c) $1\times10^{18}$ cm$^{-3}$ and (d) $1\times10^{19}$ cm$^{-3}$. (e) Effect of minority carrier lifetime and core doping on the efficiency of the heterojunction solar cell. (f) Effect of surface recombination velocity and core doping on the efficiency of a radial heterojunction solar cell for a fixed bulk minority carrier lifetime of 1 ns. (g) Effect of surface recombination velocity and core doping on the efficiency of radial heterojunction solar cell for a fixed bulk minority carrier lifetime of 50 ps. *[Figures 12(a)-(d) have been reprinted (adapted) with permission from ref [15] Copyright (2019) IEEE.]*

As mentioned earlier, the growth of radial homojunction III-V solar cells can often be complicated. Therefore, we proposed a core-shell solar cell structure which utilized a passivating carrier selective contact to form the p-n junction and achieve charge carrier separation, while simultaneously achieving surface passivation. Through both simulations and experiments, we have shown that a core-shell solar cell based on carrier selective contact is not only optically optimal but is also advantageous in terms of device performance [15]. The optical behavior, as well as a schematic of InP nanowire solar cell using electron selective contact, have already been discussed earlier in this review (see Figure 6). Now, we discuss some of the device aspects of this device. Figures 12(a)-(d) show the built-in electric field for different core doping concentrations [15]. It is quite evident that to achieve a sufficiently high built-in electric field, the p-type core doping should be larger than $10^{18}$ cm$^{-3}$. When the minority carrier lifetime is extremely low charge carrier separation must happen through electric field assisted drift instead of diffusion. Thereby, achieving very high built-in electric field is essential for high efficiency in a radial junction solar cell, especially when the minority carrier lifetime is low. Also, when the core is low or moderately doped, the junction is formed between the depleted core and p-type substrate, rather than between AZO/ZnO and p-InP, leading to an axial junction solar cell instead of radial junction solar cell [15]. Figure 12(e) further shows the effect of core doping on the efficiency of the AZO/ZnO/p-InP solar cell for different minority carrier lifetimes and it is quite evident that irrespective of the minority carrier lifetime of the core, the core has to be heavily doped to achieve sufficiently high efficiency [15]. Another important parameter that we simulated was the effect of Surface Recombination Velocity (SRV) for different core doping levels and bulk minority carrier lifetimes. Figures 12(f) and 12(g) respectively show the effect of SRV and core doping concentration on the efficiency of the solar cell when minority carrier lifetime was fixed at 1 ns and 50 ps, respectively. It is quite evident that to achieve high efficiency, AZO/ZnO/p-InP heterojunction solar cells need to have a heavily p-type doped core with a carrier concentration of the order of $1\times10^{18}$ cm$^{-3}$, while maintaining a low SRV at ZnO/p-InP interface [15]. This result is very much similar to what Li et al. [16] reported for radial GaAs homojunction. In a recent submission, we verified most of the simulated results through experimental work [76].

**Future Outlook**

The nanowire architecture allows for the fabrication of devices with new concepts which would otherwise be either too complicated or not possible to realize in planar solar cells. In the next section, we detail some of these critical concepts and discuss their importance on the future of nanowire solar cells.

*Aerotaxy*

Aerotaxy, developed at Lund University has promise towards reducing the fabrication and growth cost of III-V nanowire solar cells. Heurlin et al. claimed that the growth rate of nanowires grown using aerotaxy could be as high as 1 micron per second, which according to them is 20 to 1000 times higher than the nanowires grown using MOCVD [107]. Aerotaxy eliminates the requirement of the costly III-V single crystal substrates by facilitating the nucleation and nanowire growth on the gold agglomerate dispersed in the aerosol [107]. Therefore, aerotaxy not only reduces the cost of III-V solar cells, but it also has a high throughput compared to conventional wafer-based growth. The process of aerotaxy starts with an evaporation-condensation step for the formation of Au agglomerates. The Au agglomerates are then passed through the particle charger and differential mobility analyzer to select the appropriate size of the Au nanoparticles. After size selection, the gold nanoparticles are mixed with the GaAs precursors and then heated in a furnace for a given duration of time to grow the nanowires. For details of the technique, readers should refer to references [107-111]. The doping in aerotaxy grown nanowires is controlled through the feeding the dopants simultaneously with the growth precursors during the growth [110, 111]. Furthermore, even the growth of a heterojunction shell can be realized in by aerotaxy through feeding in the shell growth precursors after the nanowire has been grown [111]. Figure 13(a) shows a 3-D schematic of different steps in aerotaxy. Though aerotaxy has shown promise toward reducing the overall cost of nanowire solar cell, yet devices based on aerotaxy have not been able to achieve high efficiency, most probably due to the complexity in controlling the in-situ p-n junction formation [45, 110]. Nonetheless, most recently, Barrigón et al. [108] reported a single nanowire solar cell utilizing the nanowire grown using aerotaxy (see Figure 13(b)). For single nanowire, the $J_{sc}$ (220 mA/cm$^2$, normalized to the nanowire cross-section area) obtained was higher than the Schockley-Queisser limit, and the $V_{oc}$ was almost 0.6 V (see Figure 13(c)) [108]. A $V_{oc}$ of 0.6 V was not an inferior value considering they did not employ any passivation on the nanowire. Although the results achieved by Barrigón et al. [108] is remarkable, there is still a long way to go before aerotaxy can be utilized for the fabrication of nanowire solar cells on a commercial scale.

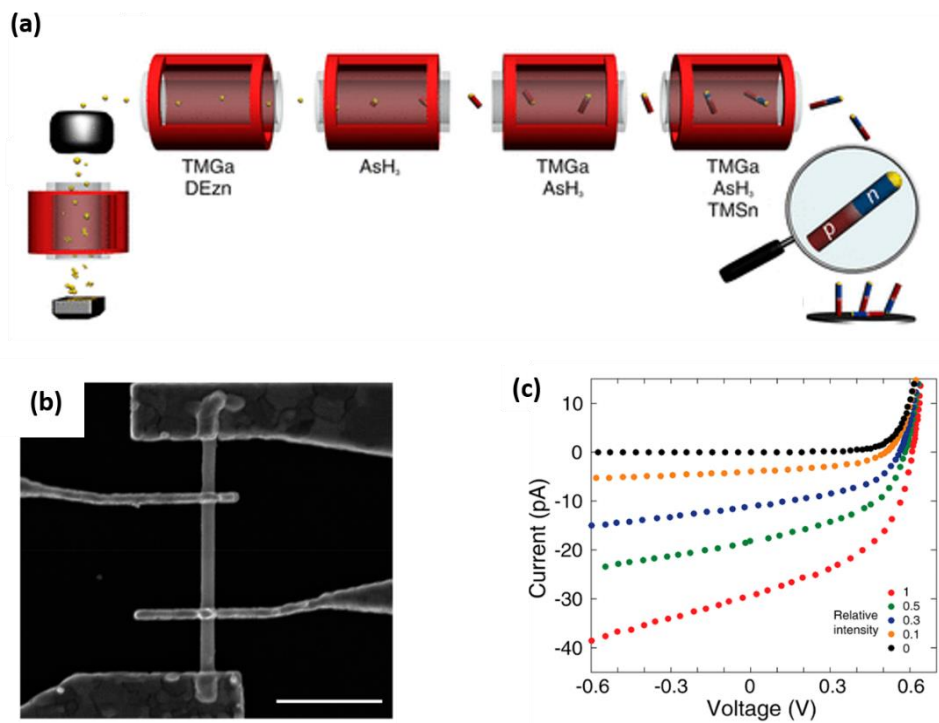

**Figure 13.** (a) Schematic showing the aerotaxy process for the synthesis of nanowire p-n junction. (b) SEM image showing a fabricated single nanowire solar cell grown by aerotaxy. (c) shows the IV curves for the single nanowire device under different laser illumination intensities. Relative intensity shows the normalized value of laser power with respect to the maximum laser power. *[Figurse 13(a)-(c) have been reprinted (adapted) with permission from ref [108] Copyright (2017) American Chemical Society.]*

*Nanowire tandem solar cells*

Another promising structure that has been proposed is III-V tandem junction solar cells on silicon or other low cost substrates. In a tandem solar cell, several absorbing materials are stacked over each other to benefit from broad bandwidth of solar radiation. In particular, the integration of III-V nanowires in tandem with silicon is one of the most promising ways to achieve low cost, high efficiency tandem junction solar cells [93, 112, 113]. In a planar solar cell, the realization of a III-V tandem junction on silicon is a relatively difficult task because of the large lattice mismatch between III-Vs and silicon which restricts the growth of III-V layers on silicon [113, 114]. However, in the case of nanowires, because of strain relaxation at the interface, researchers have been able to grow an axial tandem solar cell on silicon. For example, Heurlin et al. [115] were able to grow InP nanowires on silicon (see Figure 14(a) and 14(b)). They reported two InP nanowire solar cells axially connected to each other in series which were able to improve $V_{oc}$ by almost 67% compared to single InP nanowire. In another report, Yao et al. [113] reported the first GaAs tandem nanowire solar cells on silicon. Figures 14(c) and (d) respectively show a 3-D schematic of the device and its corresponding IV curve [113]. Another kind of tandem structure that has been proposed utilizes a radial junction for charge carrier separation [93, 116]. For example, Figure 14(e) shows a radial junction nanowire solar cell connected axially with silicon through a tunnel junction to form a tandem junction solar cell. In yet another report, Wang et al. in a simulation paper, proposed a fully radial junction nanowire tandem solar cell connected axially with each other [117] as shown in Figures 14 (f) and 14 (g). Such a device can be particularly attractive when minority carrier lifetime of nanowires is very low.

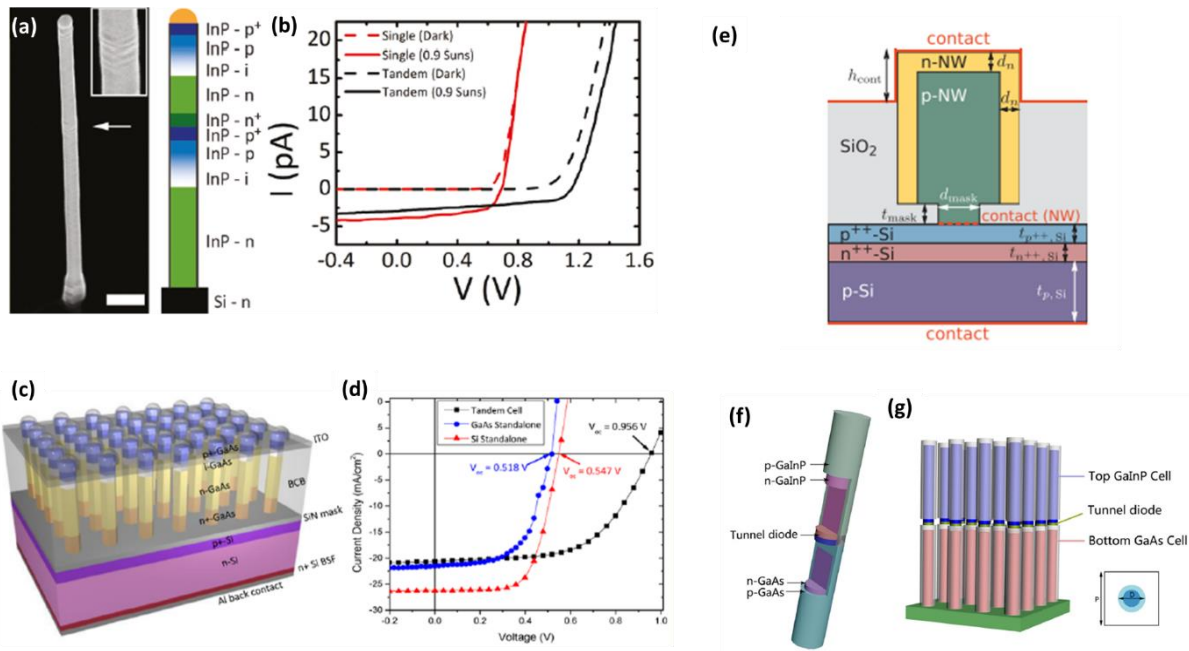

**Figure 14.** Different kinds of nanowire tandem solar cells: (a) axially connected InP nanowire grown on silicon. (b) IV curve corresponding to the solar cell shown in (a). (c) Schematic of a GaAs/Si tandem solar cell where the nanowires are grown on silicon. (d) IV curve comparing the characteristics of only GaAs nanowires, silicon only and GaAs/Si tandem solar cells. (f) and (g) show two radial junction nanowire solar cells connected axially to form a tandem solar cell. *[Figures 14(a) and 14(b) have been reprinted (adapted) with permission from ref [114]. Copyright (2011) American Chemical Society.] [Figures 14(c) and 14(d) have been reprinted (adapted) with permission from ref [115]. Copyright (2015) American Chemical Society.] [Figure 14(e) has been reprinted (adapted) with permission from ref [116]. Copyright (2018) AIP Publishing.]*

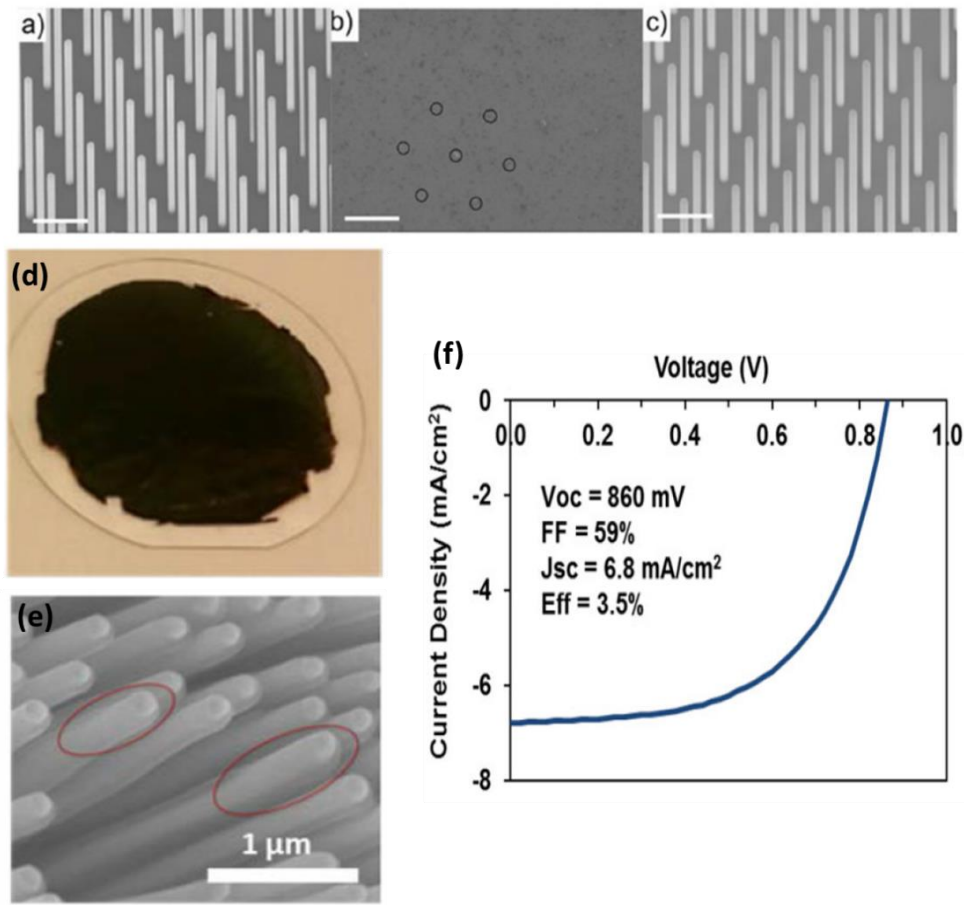

**Figure 15.** SEM images of (a) SAE grown nanowires, (b) the substrate after nanowire lift-off and (c) nanowire grown on the same substrate after three consecutive lift-off processes. (d) Nanowires transferred on a glass carrier for fabrication of solar cells. (e) SEM image showing the nanowires transferred on glass are titled leading to shunt leakage in several solar cells . (f) IV curve of the nanowire solar cell fabricated after lift-off. *[Figures 15 (a)-(f) have been reprinted (adapted) with permission from ref [129]. Copyright (2018) IEEE.]*

---

*Heteroepitaxy, lift-off and flexible solar cell*

There are several other ways in which nanowires promise to reduce the cost of III-V solar cells. For example, nanowires can be epitaxially grown on low cost substrates to reduce the cost of III-V solar cells. III-V nanowires have been grown on a variety of substrates including but not limited to graphene[118-120], carbon nanotubes [121], silicon [122], glass [123-125], ITO [126], a-Si:H [127], polycrystalline silicon[128]. However, in most cases, the quality of nanowires grown using heteroepitaxy have been found to be of lower quality compared to nanowires grown using homoepitaxy. Another approach which has the potential to reduce the overall cost of III-V photovoltaics is nanowire lift-off and substrate reuse. For example, in recent report from Lund University [129], Borgstrom et al. have shown that the group was able to grow nanowires using selective area epitaxy and reuse the substrates at least three times without compromising the quality of nanowires. Similar results have been reported by Cavalli et al.in reference [130]. Figure 15(a)-(c) respectively show the first growth of GaAs nanowire using selective area epitaxy, the substrate after nanowire lift off and nanowire growth after three lift-offs. They also reported a solar cell fabricated by transferring the nanowires to a glass carrier substrate (see Figure 15(d)). The maximum efficiency achieved was 3.5% with a $V_{oc}$ of 860 mV (see Figure 15(f)). However, nanowires transferred were tilted at a given angle, thereby making the fabrication of solar cells difficult (see Figure 15(e)) due to shunting of p-n junction in several of these cells. Another important concept that can be realized by nanowire lift-off and transfer is the fabrication of III-V/Si tandem solar cell. They showed that the III-V nanowires can be directly transferred on silicon solar cells to achieve a tandem junction [129]. When combined with low temperature fabrication processes nanowires lift-off and transfer can also lead to flexible solar cells [17].

*Spectrum splitting*

Finally, another important concept that has been proposed is multiterminal spectrum splitting, where nanowires of different materials with different bandgaps are arranged to achieve a broadband light absorption [131]. Figure 16(a) shows a 3-D schematic of III-V nanowire of three different bandgaps grown on a silicon substrate. Figure 16(b) shows a representative diagram of spectrum splitting in nanowires of different bandgap along with their expected EQEs. The proposed nanowires were InGaAs, GaAs, and AlGaAs. Dorodnyy et al. [131] further performed a detailed-balance analysis of three-terminal device shown in Figure 16(c) and claimed that an efficiency of 48.3% can be achieved from this three-terminal device.

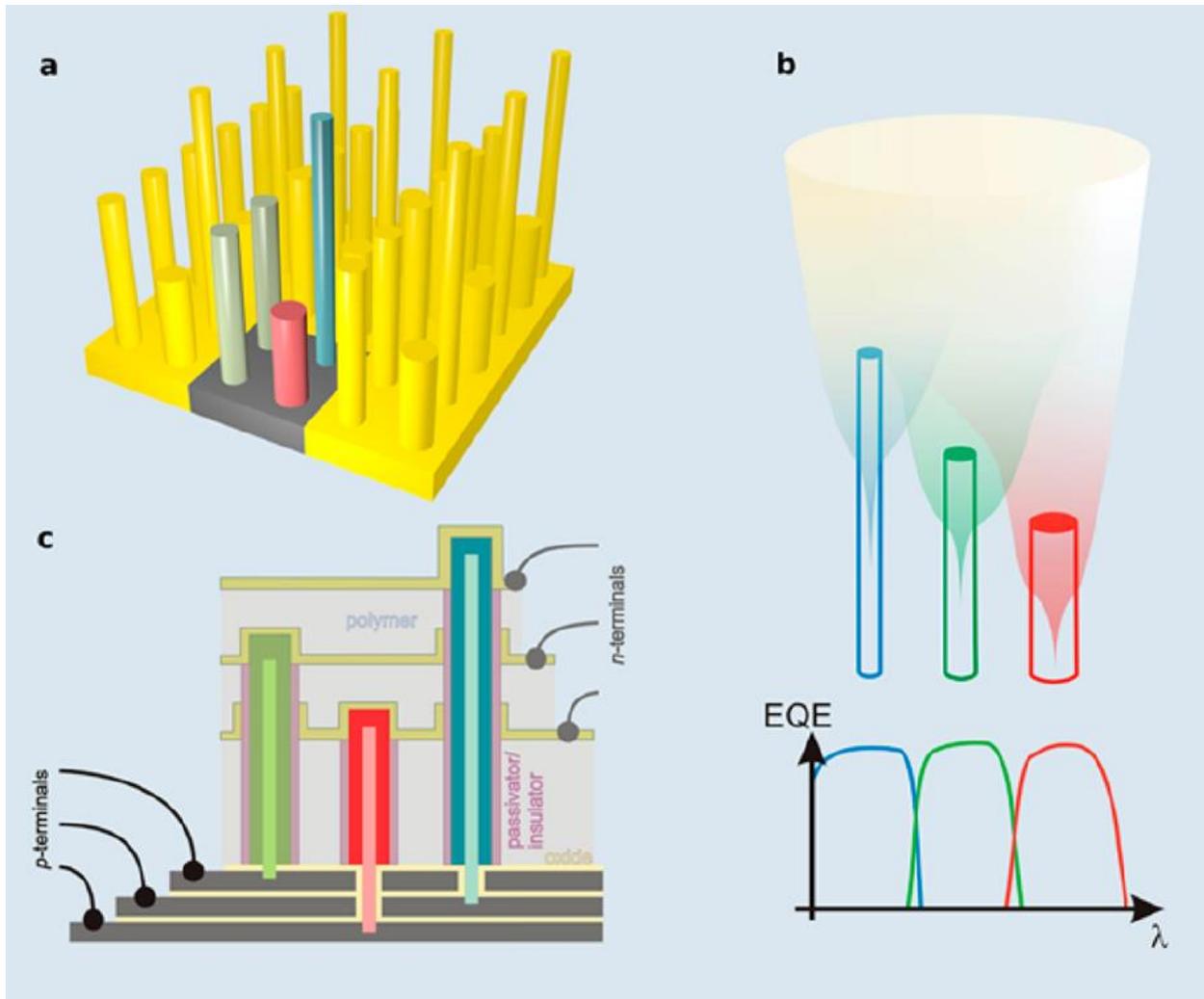

**Figure 16.** (a) 3-D schematic of an array of different sized nanowires for achieving spectrum splitting. (b) A representative diagram showing that different spectral regions of of the solar spectrum are absorbed in different nanowires of different band gaps along with matching EQEs. (c) Schematic showing how such a device can be fabricated. *[Figures 14(a)-(c) have been reprinted (adapted) with permission from ref [131]. Copyright (2015) American Chemical Society.]*

**Concluding Remarks and Challenges**

Through a range of examples that we have reviewed, it can very well be argued that nanowire architecture allows precise tuning and adjustment of both the optical and device properties of a solar cell. It also allows for the fabrication of solar cell devices, which would otherwise be either be complicated or not possible in planar junction solar cells. However, all of these excellent features of nanowire architectures come at the cost of some of the fundamental limitations of nanowire architecture, especially in terms of device behavior. Nanowire solar cells are highly susceptible to small changes in junction design, geometry, and position, which makes it harder for optimization during the growth of the nanowires. Another limitation is the large surface area, leading to high surface recombination and makes device fabrication more difficult. From the growth point of view, control of doping and p-n junction, and the growth of a shell with low surface state density are significant limitations of both axial and radial junction solar cells. New device designs such as radial heterojunctions using carrier selective contacts provide a path toward achieving high efficiency in nanowire solar cells with relatively less complexity in terms of growth and device fabrication. However, there is still a long way to go before nanowire solar cells can compete with current Si solar cells. Nevertheless, there is still great excitement in nanowire solar cells for certain niche applications, for example in self-powered micro-devices and flexible devices.

**Acknowledgment**

The authors dedicate this paper to Professor Ajoy Ghatak on the occasion of his 80th birthday. He is an exceptional teacher who has been the pillar of optics and photonics in India and contributed significantly to the field globally. The field of optics and photonics has been enriched by his contributions. The Australian Research Council is acknowledged for the financial support and the Australian National Fabrication Facility, ACT node is acknowledged for access to the facilities for some of the work referred to in this review.